\documentclass[twocolumn]{aastex631}
\usepackage[mathlines]{lineno}
\usepackage[T1]{fontenc}
\usepackage{graphicx}
\usepackage[]{natbib}
\usepackage{amsmath, multirow}
\usepackage{hyperref}
\usepackage{longtable}
\usepackage[colorinlistoftodos]{todonotes}

\newcommand{\hst}{\textit{HST}}
\newcommand{\jwst}{\textit{JWST}}

\newcommand{\eazypy}{\texttt{EAzY-Py}}

\newcommand{\scipy}{\texttt{SciPy}}

\newcommand{\lenstr}{\texttt{Lenstruction}}

\newcommand{\lya}{Lyman-$\alpha$}

\newcommand{\Heii}{He~{\sc ii}}
\newcommand{\Oii}{[O~{\sc ii}]\,$\lambda$3727}
\newcommand{\Oiii}{[O~{\sc iii}]}
\newcommand{\Oiiit}{[O~{\sc iii}]\,$\lambda$4363}
\newcommand{\OiiiUV}{O~{\sc iii}]\,$\lambda$1667}
\newcommand{\Ciii}{C~{\sc iii}]\,$\lambda$1909}

\newcommand{\Hb}{H$\beta$}
\newcommand{\Hg}{H$\gamma$}
\newcommand{\Hd}{H$\delta$}

\newcommand{\Oiiia}{[O~{\sc iii}]\,$\lambda$4959}
\newcommand{\Oiiib}{[O~{\sc iii}]\,$\lambda$5007}

\newcommand{\gal}{CANUCS-A370-z8-LAE}
\newcommand{\xiion}{$\xi_{\rm{ion}}\,$}	
\newcommand{\xiunit}{$\rm{Hz}\,\rm{erg}^{-1}$}	

\begin{document}

\title{In Search of the First Stars: An Ultra-compact and Very-low-metallicity Ly$\alpha$ Emitter Deep within the Epoch of Reionization}

\shorttitle{In Search of the First Stars}
\shortauthors{Willott et al.}

\author[0000-0002-4201-7367]{Chris J. Willott}
\affiliation{NRC Herzberg, 5071 West Saanich Rd, Victoria, BC V9E 2E7, Canada}
\email{chris.willott@nrc.ca}

\author[0000-0003-3983-5438]{Yoshihisa Asada}
\affiliation{Department of Astronomy and Physics and Institute for Computational Astrophysics, Saint Mary's University, 923 Robie Street, Halifax, Nova Scotia B3H 3C3, Canada}
\affiliation{Department of Astronomy, Kyoto University, Sakyo-ku, Kyoto 606-8502, Japan}

\author[0000-0001-9298-3523]{Kartheik G. Iyer}
\affiliation{Columbia Astrophysics Laboratory, Columbia University, 550 West 120th Street, New York, NY 10027, USA}

\author{Jon Jude\v{z}}
\affiliation{University of Ljubljana, Department of Mathematics and Physics, Jadranska ulica 19, SI-1000 Ljubljana, Slovenia}

\author[0009-0009-4388-898X]{Gregor Rihtar\v{s}i\v{c}}
\affiliation{University of Ljubljana, Department of Mathematics and Physics, Jadranska ulica 19, SI-1000 Ljubljana, Slovenia}

\author[0000-0003-3243-9969]{Nicholas S. Martis}
\affiliation{University of Ljubljana, Department of Mathematics and Physics, Jadranska ulica 19, SI-1000 Ljubljana, Slovenia}

\author[0000-0001-8830-2166]{Ghassan T. E. Sarrouh}
\affiliation{Department of Physics and Astronomy, York University, 4700 Keele St. Toronto, Ontario, M3J 1P3, Canada}

\author[0000-0001-8325-1742]{Guillaume Desprez}
\affiliation{Kapteyn Astronomical Institute, University of Groningen, P.O. Box 800, 9700AV Groningen, The Netherlands}

\author[0000-0001-9414-6382]{Anishya Harshan}
\affiliation{University of Ljubljana, Department of Mathematics and Physics, Jadranska ulica 19, SI-1000 Ljubljana, Slovenia}

\author[0000-0002-8530-9765]{Lamiya Mowla}
\affiliation{Whitin Observatory, Department of Physics and Astronomy, Wellesley College, 106 Central Street, Wellesley, MA 02481, USA}

\author{Ga\"el Noirot}
\affiliation{Space Telescope Science Institute, 3700 San Martin Drive, Baltimore, Maryland 21218, USA}

\author{Giordano Felicioni}
\affiliation{University of Ljubljana, Department of Mathematics and Physics, Jadranska ulica 19, SI-1000 Ljubljana, Slovenia}

\author[0000-0001-5984-0395]{Maru\v{s}a Brada{\v c}}
\affiliation{University of Ljubljana, Department of Mathematics and Physics, Jadranska ulica 19, SI-1000 Ljubljana, Slovenia}
\affiliation{Department of Physics and Astronomy, University of California Davis, 1 Shields Avenue, Davis, CA 95616, USA}

\author[0000-0003-2680-005X]{Gabe Brammer}
\affiliation{Cosmic Dawn Center (DAWN), Denmark}
\affiliation{Niels Bohr Institute, University of Copenhagen, Jagtvej 128, DK-2200 Copenhagen N, Denmark}

\author[0000-0002-9330-9108]{Adam Muzzin}
\affiliation{Department of Physics and Astronomy, York University, 4700 Keele St. Toronto, Ontario, M3J 1P3, Canada}

\author[0000-0002-7712-7857]{Marcin Sawicki}
\affiliation{Department of Astronomy and Physics and Institute for Computational Astrophysics, Saint Mary's University, 923 Robie Street, Halifax, Nova Scotia B3H 3C3, Canada}

\author[0000-0002-0243-6575]{Jacqueline Antwi-Danso}
\affiliation{David A. Dunlap Department of Astronomy and Astrophysics, University of Toronto, 50 St. George Street, Toronto, Ontario, M5S 3H4, Canada}
\altaffiliation{Banting Postdoctoral Fellow}

\author[0000-0002-5694-6124]{Vladan Markov}
\affiliation{University of Ljubljana, Department of Mathematics and Physics, Jadranska ulica 19, SI-1000 Ljubljana, Slovenia}

\author[0000-0002-9909-3491]{Roberta Tripodi}
\affiliation{University of Ljubljana, Department of Mathematics and Physics, Jadranska ulica 19, SI-1000 Ljubljana, Slovenia}
\affiliation{INAF - Osservatorio Astronomico di Roma, Via Frascati 33, Monte Porzio Catone, 00078, Italy}


\begin{abstract}
We present {\it JWST} observations of a gravitationally-lensed, extremely metal-poor galaxy at redshift $z=8.203\pm 0.001$  from the CANUCS survey. Based on the low oxygen to Balmer line ratios we infer a gas-phase metallicity of $12+{\rm log(O/H)}=6.85$ (1.4\% solar), making CANUCS-A370-z8-LAE one of the most metal-poor galaxies known at $z>7$. The galaxy has a high H$\beta$ equivalent width of $225\pm50$\,\AA, small half-light radius of only $r_{\rm hl} = 38 ^{+3}_{-19} $\,pc, and high star-formation-rate density of $50 - 100\,M_{\sun}$\,yr$^{-1}$\,kpc$^{-2}$. The galaxy shows high equivalent width Lyman-$\alpha$ emission with an inferred Lyman-$\alpha$ escape fraction of $0.21 \pm 0.05$. The high escape fraction of Lyman-$\alpha$ is likely due to the compact starbursting nature of the galaxy combined with its location in an overdensity traced by at least two other galaxies spectroscopically confirmed to lie within $\delta z = 0.01$ that have helped to reionize the environment. The low metallicity of CANUCS-A370-z8-LAE is best explained by a model where infalling metal-poor gas dilutes the interstellar medium, rather than being a young galaxy forming its first stellar populations.
\end{abstract}


\section{Introduction}
\label{sec:intro}

A key goal for the {\it James Webb Space Telescope} (\jwst, \citealt{Gardner2023}) is to identify the first populations of stars that arose from pristine gas, devoid of heavy elements, the so-called Population III stars. The first stars are expected to be extremely massive with very short lifetimes that end in violent supernovae \citep{Bromm1999}. Consequently, galaxies rapidly become enriched in heavy elements, and finding galaxies with a high star formation rate (SFR) and no metals will be difficult \citep{Katz2023}. The best prospects for detecting Population III objects may therefore be from the first supernovae strongly magnified by gravitational lensing \citep{Windhorst2018, Venditti2024} or in very low mass regions discovered adjacent to more enriched galaxies \citep{Rydberg2013, Zackrisson2015}.

Spectroscopic observations with \jwst\ have allowed the measurement of metallicities for hundreds of high-redshift galaxies. However, rather surprisingly, only shallow evolution toward lower metallicities at high redshifts has been observed \citep{Nakajima2023, Curti2024, Sanders2024, Langeroodi2024arXiv}. The typical metallicity of galaxies at redshifts 7 to 9 (corresponding to 750 to 500 million years after the Big Bang) is 10\% times the solar abundance, with almost all galaxies in this redshift range having metallicity in the range 3\% to 30\% solar, and a well-developed mass-metallicity relation already in place \citep{Nakajima2023, Curti2024}. This may be indicative of a {\it metallicity floor} due to the aforementioned rapid enrichment provided by the first very massive stars \citep{Wise2012, Nishigaki2023}.

The lowest metallicity reionization-era object measured so far is the $z=6.6$ \lya\ emitter LAP-1 \citep{Vanzella2023}. This strongly-lensed object is likely a single star cluster with very low mass of only about $10,000\,M_{\sun}$ and metallicity estimated to be less than 0.4\% solar. \cite{Fujimoto2025} additionally report a lensed $z=6.5$ low-mass galaxy that may have metallicity less than 1\% solar based on model fitting to the photometry. This may indicate that to find the first stars one has to look at strongly-lensed galaxies that have low stellar masses and where individual star-forming regions with different metallicities may be separated. 

In the local Universe, low-metallicity, low-mass galaxies are often very strong emitters of \lya\ photons \citep{Izotov2024} and Lyman continuum ionizing radiation \citep{Annibali2022}. High-redshift analogs of these galaxies may therefore be important contributors to the process of cosmic reionization. Strong Lyman-$\alpha$ emission is, however, rarely observed at redshifts $z>7$ due to absorption by intervening neutral hydrogen in the intergalactic medium (IGM; \citealt{Stark2010, Tang2024, Jones2025}). When high \lya\ escape fractions are observed, calculation of the ionized bubble size to allow transmission usually requires UV ionizing continuum from multiple galaxies in an overdensity \citep{Saxena2023, Jung2024, Chen2024, Witstok2025_3LAEs}. Therefore, \lya\ transmission at this epoch often requires favorable conditions on both the galaxy-scale and protocluster scale (e.g. \citealt{Witten2024}).

In this paper, we discuss the galaxy \gal, which we have identified from NIRCam imaging and follow-up NIRSpec spectroscopy to be a very compact and strongly-lensed galaxy displaying both \lya\ emission and very low metallicity nebular gas at a redshift of $z=8.203\pm 0.001$. This galaxy was first identified from Hubble Frontier Fields imaging \citep{Lotz2017} in Abell 370. In \cite{Ishigaki2018}, it is object HFF6C-5187-5411 with $z_{\rm phot} =7.38_{-5.89}^{+0.82}$ and lensing magnification of 17.57. \cite{Yang2022} adopted $z_{\rm phot} =7.70$ and fit a source-plane effective radius of $r_e=0.07 \pm 0.03$ kpc -- one of the smallest $z\approx 8$ galaxies in their sample.

In Section 2 we present the data used in this work. Section 3 describes the physical inferences made possible by these data. Section 4 discusses the nature of this galaxy and its implications. In Section 5 we give our conclusions. We assume a flat $\Lambda$CMD cosmology, with $\Omega_\Lambda=0.7$, $\Omega_{\rm m}=0.3$, and $H_0=70~{\rm km\,s^{-1}\,Mpc^{-1}}$. Magnitudes are given in the AB system \citep{Oke1983}.  

\begin{figure*}
    \centering
    \includegraphics[width=\linewidth]{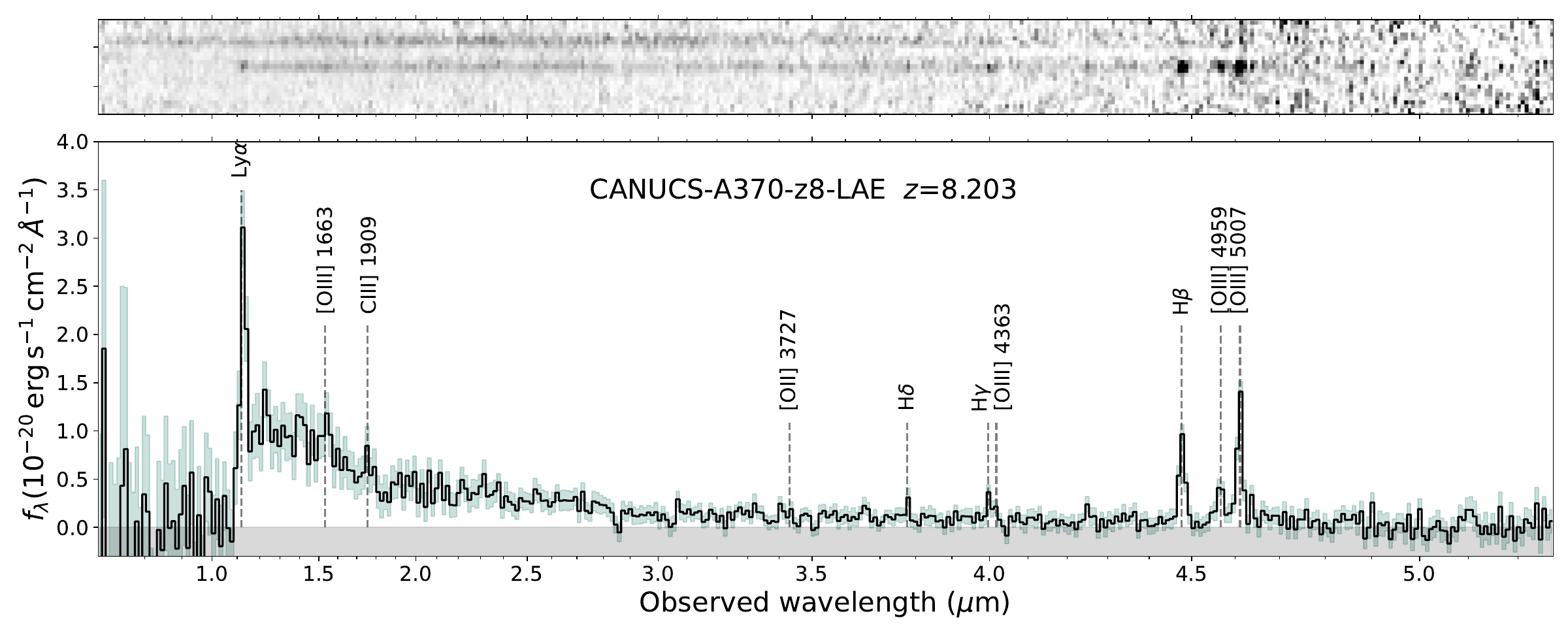}
    \caption{NIRSpec prism spectrum of \gal. The upper panel shows the 2D spectrum of the target in the central row with a lower redshift galaxy spectrum near the top of the panel. Several emission lines, including \lya, are clearly visible in the 2D spectrum. The lower panel shows a 1D optimal extraction. Expected strong emission lines are labeled. As well as strong \lya, this spectrum is notable for the very high ratios of \Hb\ to \Oiiib\ and \Oiiib\ to \Oii, indicative of a very low gas-phase metallicity.}
    \label{fig:spectrum}
\end{figure*}

\section{Data}
\label{sec:data}

\subsection{Imaging}
\label{sec:imaging}

We observed \gal\ in imaging mode with the NIRCam \citep{Rieke2023} and NIRISS \citep{Doyon2023} instruments as part of the JWST Cycle 1 program CANUCS \citep{Willott2022} in December 2022. NIRCam observations were made in the F090W, F115W, F150W, F200W, F277W, F356W, F410M, and F444W filters for $6.4\,{\rm ks}$ each. With NIRISS the F115W, F150W, and F200W filters were used for $2.3\,{\rm ks}$ each. The data were processed along with the existing \hst\ imaging as described in earlier CANUCS papers (\citealt{Desprez2024, Willott2024}, see also \citealt{Sarrouh2025arXiv}). Mosaic images are generated by drizzling onto pixel scales of 40 milli-arcsec for source detection and photometry. The NIRCam short-wave (SW) images are also drizzled onto a 20 milli-arcsec scale for morphological analysis. From the resulting photometric catalog, this galaxy has an \eazypy\ \citep{Brammer2008} photometric redshift of $z_{\rm phot} = 7.93^{+0.17}_{-0.06}$. This photometric redshift fitting followed the method of \cite{Asada2025} using templates with strong emission lines and increased neutral hydrogen absorption at $z>7$.

\subsection{Spectroscopy}
\label{sec:spectra}

The Abell 370 field was observed on 5 September 2023 with NIRSpec low-resolution prism follow-up using the Micro-Shutter Assembly (MSA; \citealt{Ferruit2022}) as part of the CANUCS program. Strongly-lensed $z>7$ galaxies were targeted with high priority to study the stellar populations and interstellar medium (ISM) conditions in low-mass galaxies within the Epoch of Reionization. Three MSA configurations were observed to enable the large gaps between the four MSA quadrants to be dithered across. Each configuration was observed for 2.9\,ks with the targets nodded along a 3-shutter slitlet. \gal\ was included in all three MSA configurations, so the total exposure time was 8.7\,ks.  

Details of the NIRSpec processing are given in \cite{Desprez2024}. For this galaxy, we use a version of the {\tt msaexp} package \citep{Brammer2022zndo} with master background subtraction to avoid oversubtraction due to a nearby galaxy along the slitlet in some nod positions. The wavelength solution accounts for the intra-shutter location of the target. A one-dimensional (1D) optimal extraction is performed from the combined two-dimensional (2D) spectrum. A slit-loss correction is applied based on a polynomial fit to the NIRCam photometry. Flux uncertainties in NIRSpec prism spectra from the pipeline are known to be underestimated due to inter-pixel correlations and non-Gaussian noise \citep{ArrabalHaro2023b}. We test for this by extracting line fluxes and errors at random positions in the 1D spectrum. We find a factor of 1.5 increase in uncertainty is required to bring the number of false identifications at $>2\sigma$ in line with expectations, so we multiply the error array by this value.

The 2D and 1D spectra of \gal\ are shown in Figure\,\ref{fig:spectrum}. Strong emission lines typical of galaxies at these redshifts such as \Oiii, \Hb\ and \Hg\ are visible. The ratio of \Hb\ to \Oiiib\ is unusually high. The \Oii\ line is not detected, so the high  \Oiiib\ to \Oii\ ratio indicates a high ionization parameter. Together, these results indicate a very low oxygen abundance in the nebular gas (e.g. \citealt{Nakajima2022}), as will be discussed in Section \ref{sec:metal}. 

Another unusual aspect of this spectrum is the appearance of strong \lya\ emission. Strong \lya\ emission is only visible in the NIRSpec low-resolution prism spectra for 10\% of $z\approx 8$ galaxies \citep{Jones2025}. Given the highly neutral state of the IGM at this epoch \citep{Planck2020} this suggests that \gal\ resides within a large ionized bubble that allows \lya\ radiation to escape. 

\section{Measurements}
\label{sec:measure}

\subsection{Spectral Fitting}
\label{sec:specfit}

\begin{figure*}
    \centering
    \includegraphics[width=0.49\linewidth]{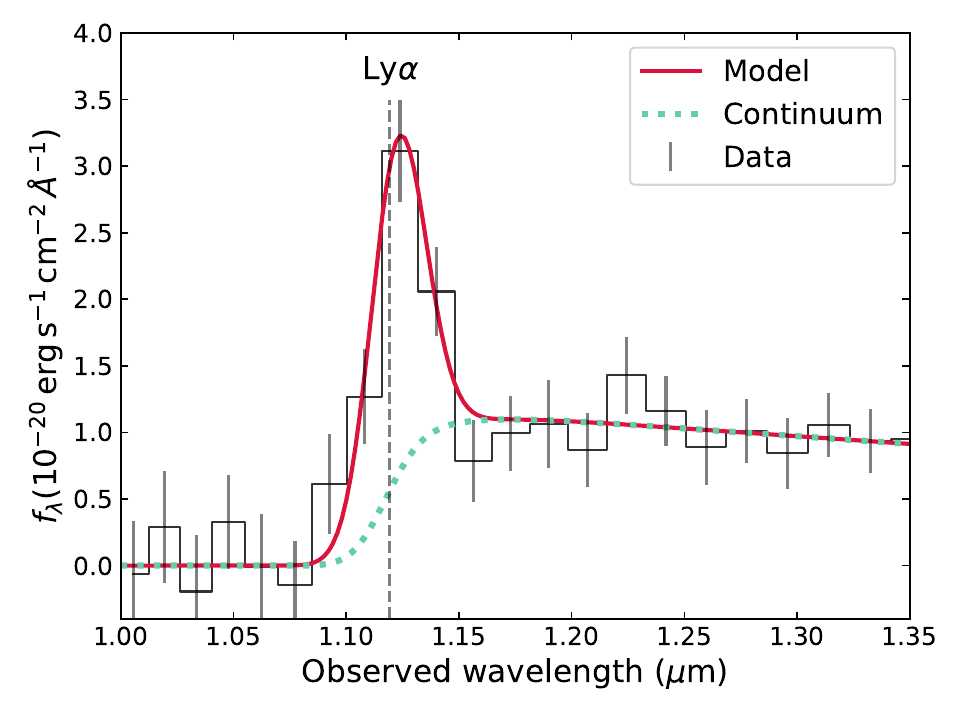}
    \includegraphics[width=0.49\linewidth]{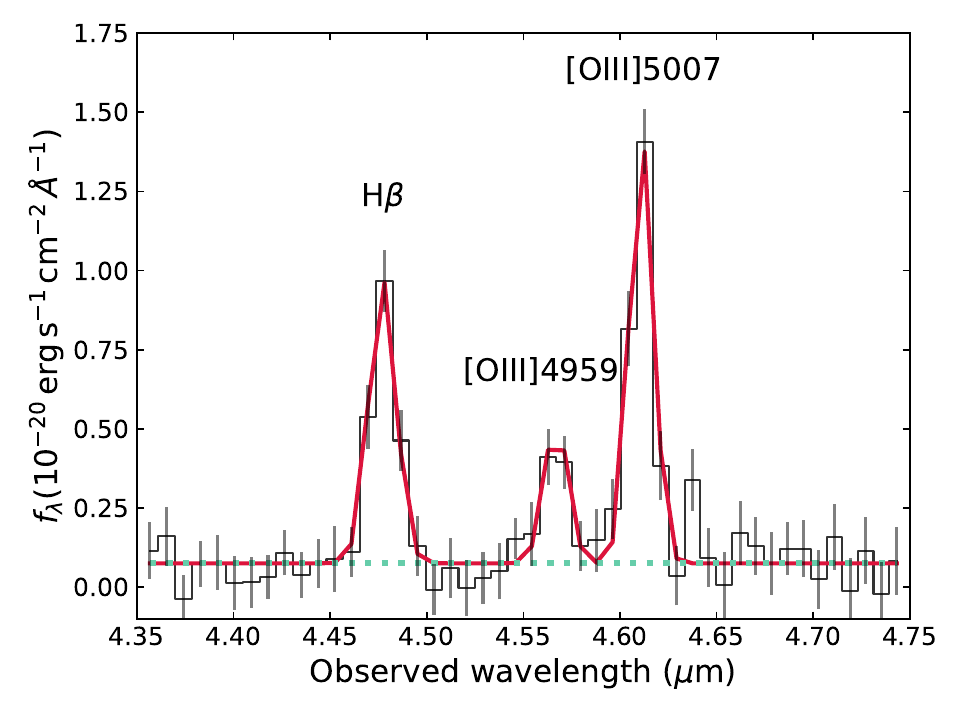}
    \caption{Emission line fits (red lines) for \gal\ compared to the data (black steps with $1\sigma$ uncertainties). Left: The \lya\ line is modeled as a Gaussian at the systemic redshift (marked by the dashed vertical line) added to an underlying $\beta=-1.8$ UV continuum. This model is subjected to absorption from a fully neutral IGM with the galaxy at the center of a 2\,pMpc radius ionized bubble. The full model is convolved with the NIRSpec prism line spread function and then rebinned to the observed spectral resolution before fitting. The continuum contribution is shown as a green dotted line. Right: Gaussian fits to the observed \Hb\ and \Oiii\ lines show very strong \Hb\ emission that is not broader than the \Oiii\ emission, arguing against a broad-line AGN explanation for the strong \Hb\ emission.}
    \label{fig:linefits}
\end{figure*}

We fit spectral regions near emission lines using the Levenberg-Marquardt non-linear least-squares algorithm. This method returns uncertainties that account for the covariance between parameters. Results of fitting the 1D spectrum are shown in Figure \ref{fig:linefits}. The \lya\ spectral regions of galaxies at this epoch are complex owing to \lya\ absorption from nearby regions that can have a range of density and neutral fraction combined with \lya\ emission that is resonantly scattered. Given the moderate S/N of our spectrum we are unable to simultaneously constrain the properties of the IGM, an adjacent damped \lya\ absorber and an ionized bubble. From the presence of strong \lya\ emission we expect minimal effect from any proximate damped \lya\ absorber (although examples of such cases do exist and require the \lya\ emission to be offset from the continuum, e.g. \citealt{Witstok2025z13LAE}). However, in the case of \gal\ there is no strong downturn in the UV continuum near \lya\ as would be expected for a high column density absorber. 

We set the intrinsic \lya\ emission to be a Gaussian at the systemic redshift determined via fitting the rest-optical lines. Any velocity offset within the galaxy would be much less than one pixel at this wavelength ($4000\, {\rm km\,s}^{-1}$) so can be neglected. We model the continuum as a power-law with $\beta=-1.8\pm 0.15$ (where $f_{\lambda} \propto  \lambda^{\beta}$) based on a fit to the NIRCam photometry at rest-frame 1500 to 3400\,\AA. As mentioned previously, the spectral S/N is not high enough to constrain the IGM neutral fraction, $x_{\rm HI}$, and fully-ionized bubble size, $R_{\rm ion}$, but setting $x_{\rm HI}=1$ and $R_{\rm ion}=2\,{\rm pMpc}$ allows a good fit of a combined line plus continuum model to the data. This model is generated at higher spectral resolution, convolved with the NIRSpec prism line spread function (including a correction factor of 1.6 due to this compact source underfilling the shutter as in \citealt{deGraaff2024}) and rebinned to the pixel sampling of the data. The parameters of the continuum amplitude, \lya\ flux and \lya\ Gaussian dispersion are left as free parameters. From the fit we calculate the \lya\ rest-frame equivalent width to be EW(Ly$\alpha)=63 \pm 9$\,\AA. The reduced $\chi^2$ is 0.62, indicating a good fit and that the flux uncertainties may be a little too high in this region of the spectrum. Given that reionization is expected to have begun by this redshift and consequently the IGM is not completely neutral \citep{Planck2020, Kageura2025arXiv}, we also tested a fit assuming $x_{\rm HI}=0.7$ and achieved a similar goodness of fit with only a 3\% change in the intrinsic \lya\ flux.

The rest-frame optical lines of \Hb\ and \Oiii\ are plotted in the right panel of Figure \ref{fig:linefits}. All three lines are simultaneously fit by observed Gaussians at a single redshift. In this case, we do not fit a higher resolution model convolved with the line spread function. The ratios of \Oiiia\ and \Oiiib\ are constrained by their known ratio of 2.98 from atomic physics. The free parameters are the redshift, a constant continuum level, Gaussian line amplitudes, and observed dispersions of \Oiii\ and \Hb. The redshift resulting from this fit is $z=8.203\pm 0.001$. We find an observed Gaussian dispersion of $69 \pm 8$\,\AA\ for \Hb\ and $61 \pm 6$\,\AA\ for \Oiii. Based on the prism dispersion change with wavelength, we would expect the \Hb\ dispersion to be 7\% greater than that of \Oiiib. We find it to be 11\% greater, but the difference is well within the $1\sigma$ uncertainties. Since the FWHMs of these lines are consistent, we find no evidence for an AGN broad-line region in \gal. The \Hb\ rest-frame equivalent width is EW(H$\beta)=225\pm50$\,\AA\ placing this galaxy in the high EW(H$\beta$) class of galaxies according to \cite{Nakajima2022}. 

We fit other important emission lines in a similar manner. \Hg\ and \Oiiit\ are partially blended at the spectral resolution of the prism. However, the fluxes of both lines can often be well constrained at $z>8$ with high enough S/N data, due to the increase in spectral resolution toward the red end of the prism wavelength range (e.g. \citealt{Hsiao2024, Mowla2024, Tripodi2024arxiv}). Given the low oxygen abundance and faintness of this galaxy, \Oiiit\ is not significantly detected (we adopt a $2\sigma$ threshold for significant detections) and are unable to place a strong limit on the ratio of  \Oiiit\ to \Oiiib\ and hence the electron temperature. 
There is a peak in the 1D spectrum at the expected position of \OiiiUV, but given the low significance and possible blending with \Heii\ we do not fit that line. Also, in the UV, there is a single pixel spike at the expected wavelength of \Ciii, but further data would be required to confirm a significant detection.

The unblended \Hd\ line is fit in order to determine a Balmer decrement in combination with \Hb. We find a ratio of \Hd\ to \Hb\  of $0.19 \pm 0.08$.
This is within $1\sigma$ of the value of 0.25 that is expected for Case B with no dust reddening. Considering also the presence of strong \lya\ emission in the observed spectrum, we assume no dust reddening of the nebular gas.

The low ionization \Oii\ line is not significantly detected. We determine a $2\sigma$ upper limit from fitting the spectrum at the expected position and use this to derive a lower limit on the ratio of \Oiiib\ to \Oii\ of ${\rm O32}>8.8$, indicating very high ionization, consistent with the known correlation between O32 and EW(H$\beta$) \citep{Nakajima2022}. The results of emission line fitting are given in Table \ref{tab:galprop}.

\subsection{Lensing Modeling}
\label{sec:lensmod}

We use the CANUCS gravitational lensing model of the Abell 370 cluster \citep{Gledhill2024} to derive the magnification and enable a source-plane morphology analysis. The lensing model is not highly constrained in this region due to a lack of multiple images nearby and a massive elongated mass component accounting for the southern overdensity of cluster galaxies.  The best-fit lensing magnification at the location and redshift of \gal\ is $\mu=8.0$. The uncertainty in the lensing model is derived from 100 randomly drawn lens models from the Bayesian sample of the Monte Carlo Markov Chain run (see \citealt{Gledhill2024} for details). The distribution of acceptable models in these 100 random draws has a wide range from $9.6 < \mu < 21.8$ at 68\% confidence. We note that the best-fit model is not contained within the $1\,\sigma$ range, as the likelihood distribution is not symmetric, with the lowest $\chi^2$ values tending to lie at the low end of the magnification range, rather than near the median. To be conservative, we adopt the best-fit magnification $\mu=8.0$ as our default. However, for the morphology analysis in the next section we will additionally use all 100 iterations of the model to include the lensing uncertainty.


\subsection{Morphology Fitting}
\label{sec:size}

\gal\ is extremely compact in NIRCam imaging, despite being strongly lensed with $\mu \approx 8$. The lensing field at this location is highly tangential, with the tangential component of the magnification equal to $\approx 7$. \gal\ appears marginally resolved along this direction in the NIRCam SW images in the F150W and F200W filters. It does not appear resolved in the radial direction, consistent with a small intrinsic size. 

We derive a size estimate for the galaxy by fitting the F200W image which has the highest S/N of the SW filters, noting this corresponds to rest-frame UV emission at $\approx 2200$\,\AA. We use the F200W mosaic image that has been drizzled onto a pixel scale of 20\,milli-arcsec (c.f. the native pixel scale is 33\,milli-arcsec). We use \lenstr\ \citep{Birrer2015,Birrer2018,Yang2020} which is a source reconstruction tool based on forward modeling the appearance of the source in the image plane accounting for the distortion by lensing and the instrumental point spread function (PSF). The surface brightness distribution in the source plane is approximated by a two-dimensional elliptical S\'{e}rsic profile. \lenstr\ employs a Bayesian inference formalism to estimate the full posterior distribution of all source model parameters. We use a 20\,milli-arcsec empirical PSF generated using the procedure described in \citet{Sarrouh2025arXiv}. Briefly, this procedure identifies unsaturated stars in the field, then those stars are shifted, masked, normalized, and median-stacked. Stars whose radial profiles are outliers are iteratively removed from the stack. The outer region of the empirical PSF (radius $>0.5$ arcsec) has low S/N, so is replaced by a custom {\texttt{WebbPSF}} model \citep{Perrin2014}, built using the date and orientation of observation.

\begin{figure}
    \centering
    \includegraphics[width=\linewidth]{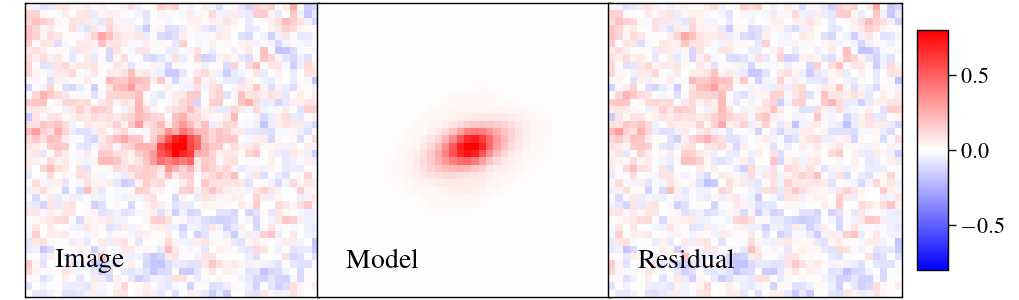}
    \caption{Left: NIRCam F200W image of \gal\ on a 20\,milli-arcsec pixel scale in units of nJy per pixel. The field shown is 0.8\,arcsec. The galaxy appears slightly elongated at a position angle close to the lensing shear direction  of $108^\circ$. Middle: \lenstr\ model reconstruction of a zero ellipticity S\'{e}rsic $n=1$ model with half-light radius = 8\,milliarcsec. This reconstruction is for the best fit lensing model ($\mu=8.0$). Right: residual image showing that this model fully represents the flux distribution of \gal. The small excess of positive flux in the north-east quarter of the image may be due to extended emission from \gal\ or, more likely, a foreground galaxy to the northeast that lies just outside the cutout.}
    \label{fig:lenstruction}
\end{figure}

Fitting lensed galaxies with a highly tangential deflection field leads to strong size constraints along one axis only. The resulting best-fit shapes are often elongated perpendicular to the shear due to this lack of constraint, and that is what we find when we leave the shape unconstrained. We therefore assume that the source-plane shape has zero ellipticity to estimate the most likely size along both axes. Under this assumption, and fixing the S\'{e}rsic index to $n=1$, we measure an intrinsic (before lensing) half-light radius of $r_{\rm hl} = 8.0 ^{+0.7}_{-4.0} $\,milliarcsec. This uncertainty range was derived using a set of 100 randomly drawn lens models that provide acceptable fits to account for both the imaging and lensing uncertainties. A fit allowing the ellipticity to vary produces a best fit with $e=0.50$ and similar circularized half-light radius of 7 milliarcsec. A comparison of the residuals for the zero and free ellipticity cases shows no significant improvement when adding these two extra free parameters. Therefore, we assume the zero ellipticity case. For such a compact size we are unable to constrain the S\'{e}rsic index of the source, so we fix it to $n=1$. If we use S\'{e}rsic indices of 0.5 and 2, the resulting half-light radii lie well within the reported uncertainty. Note that the relatively large uncertainty on $r_{\rm hl}$ is a consequence of the uncertainty in the lens model and magnification. As discussed in Section \ref{sec:lensmod}, the best-fit lens model has relatively low magnification, but the distribution has a tail to higher magnifications. For the size, this results in a tail in the uncertainty distribution toward smaller sizes. The F200W image, best-fit model, and residuals are shown in Figure\,\ref{fig:lenstruction}.

Using the angular scale at this redshift of 4.74\,kpc per arcsec, we determine the physical half-light radius to be $r_{\rm hl} = 38 ^{+3}_{-19} $\,pc. This is somewhat smaller than the $r_{\rm hl} = 70 \pm 30$ pc determined by \cite{Yang2022} using \hst\ imaging with poorer spatial resolution. Our size measurement shows \gal\ to be an extremely compact galaxy.

As an additional check on this size measurement, we measure the F200W size of the source in the observed plane by fitting a S\'{e}rsic profile with {\texttt{Galfit}} \citep{Peng2010}. We obtain an apparent half-light radius of $0.5 \pm 0.3$\,kpc. Dividing by the best-fit tangential magnification factor of 7 (because the source is strongly sheared in this direction) gives a source-plane half-light radius of $70 \pm 40$\,pc. This is consistent with the best-fit \lenstr\ radius calculated assuming the same best-fit lensing model.

\begin{figure*}
    \centering
    \includegraphics[width=0.49\linewidth]{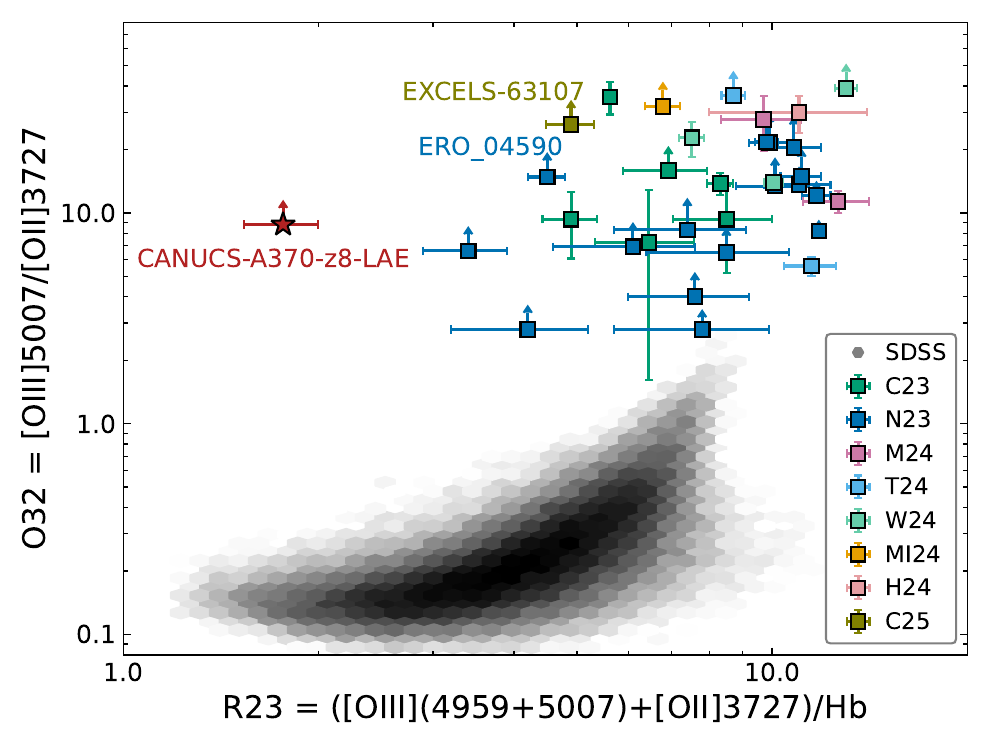}
    \includegraphics[width=0.5\linewidth]{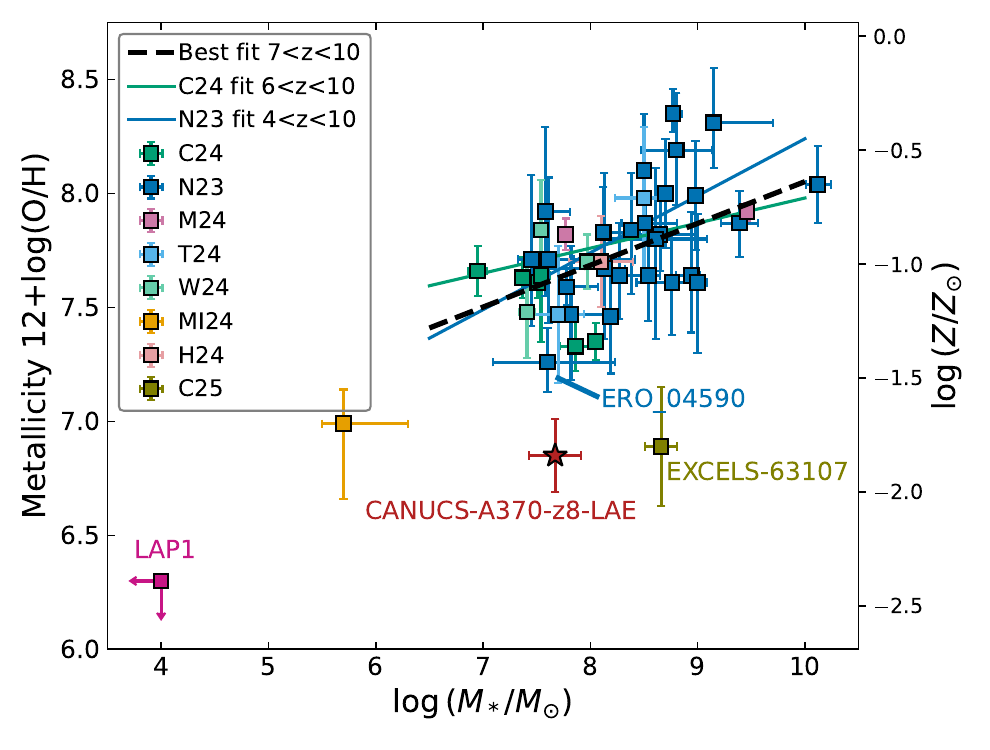}
    \caption{Left: strong emission line ratio diagnostic for redshift $7<z<10$ galaxies with JWST measurements (colored symbols) and SDSS local galaxies (gray density hexmap). \gal\ has the lowest R23 ratio of known $z>7$ galaxies, indicating a very low oxygen abundance. The JWST sample is from JADES (DR1 galaxies, C23: \citealt{Cameron2023}; LAEs, W24: \citealt{Witstok2025_3LAEs}), CEERS, GLASS and SMACS0723-ERO (N23: \citealt{Nakajima2023}), MACS1149 (M24: \citealt{Morishita2024auroral}), Abell1703 (T24: \citealt{Topping2025}), CANUCS Firefly (MI24: \citealt{Mowla2024}), MACS0647 JD1 (H24: \citealt{Hsiao2024}) and EXCELS-63107 (C25: \citealt{Cullen2025}). Right: gas-phase metallicity vs. stellar mass for a similar sample of $7<z<10$ galaxies. The JADES data, labeled C24, are from \citealt{Curti2024} in this plot. Also shown is the very low metallicity lensed star cluster, LAP-1 at $z=6.6$ \citep{Vanzella2023}. The mass and metallicity for the Firefly is plotted for a single star cluster (FF-6), since that was the only cluster with measured metallicity (MI24: \citealt{Mowla2024}). The best fit for the galaxy sample is shown by the black dashed line. This is similar to the best fit relations of \cite{Curti2024} and \cite{Nakajima2023}. \gal\ is a 4.7$\sigma$ outlier relative to the dispersion of other galaxies around the best fit line, with metallicity well below that of other $7<z<10$ galaxies.}
    \label{fig:metal}
\end{figure*}

\subsection{Spectral Energy Distribution Fitting}
\label{sec:bagpipes}
The spectral energy distribution (SED) fitting code {\texttt{Bagpipes}} \citep{Carnall2018} was used to fit the \jwst\ imaging data to derive galaxy physical parameters for \gal. PSF-homogenized photometry in a 0.3" diameter aperture was used to maximize S/N. To avoid issues with correctly fitting the \lya\ break and emission line region, only wavelengths of 1.5\,$\mu$m and longer were fit (filters F150W, F200W, F277W, F356W, F410M, F444W). The redshift is fixed at the spectroscopic redshift. The nebular ionizing parameter was set to ${\rm logU} = -1.5$ motivated by the high observed O32 ratio \citep{Papovich2022}. The models use a \cite{Kroupa2001} initial mass function.
A double power-law star formation history was adopted, but the resulting best fit shows only a steeply rising star formation history, as expected for a galaxy with very high EW(H$\beta$) and no evidence of a Balmer break in the spectrum or in photometry. Resulting physical parameters from the {\texttt{Bagpipes}} fit are given in Table \ref{tab:galprop}.

Posterior values of stellar mass and SFR were adjusted for an aperture correction from 0.3" to total  (Kron) flux of factor 2.3 and for the `best' gravitational lensing magnification of $\mu=8.0$ described in Section \ref{sec:lensmod}. The stellar mass is fairly well constrained for this galaxy (modulo the lensing uncertainty described previously). At this redshift range, the stellar mass can be uncertain for galaxies without spectroscopy because strong emission lines can account for large and uncertain fractions of the F444W flux (e.g. \citealt{Desprez2024} find fractions of 0.5 are common). This galaxy has a moderate combined rest-frame equivalent width of \Oiii\ and \Hb\ of 620\,\AA\ as measured with NIRSpec, which compares well with the {\texttt{Bagpipes}} model best fit of 720\,\AA. In addition, the F410M filter is free of strong emission lines, so the rest-frame optical continuum and hence stellar mass are well constrained. The star formation rate from the {\texttt{Bagpipes}} fit of $SFR_{\rm BAGPIPES} = 0.54\,M_{\sun}$ yr$^{-1}$ is quite similar to the star formation rate from the \Hb\ line luminosity of $SFR_{\rm H\beta} = 0.44\,M_{\sun}$ yr$^{-1}$, calculated assuming the relationship of \cite{Reddy2022} for galaxies with low metallicity. 

\section{Results}
\label{sec:results}

\subsection{Metallicity}
\label{sec:metal}

We determine the gas-phase metallicity of \gal\ using the strong optical emission lines of \Hb, \Oiii\ and the nondetection of \Oii. We apply several different calibrations based on local galaxies and distant galaxies observed with \jwst. The R23 ratio, ([OIII](4959+5007)+[OII]3727)/Hb, has been found to be the best strong line ratio for low-metallicity galaxies at high-redshift \citep{Laseter2024,Langeroodi2024arXiv}. When combined with O32 ([OIII]5007/[OII]3727), the two branches of low- and high-metallicity are easily distinguished. This is illustrated in the left panel of Figure \ref{fig:metal} where O32 is plotted against R23. The local SDSS galaxies with low R23 have low O32 and consequently have high metallicities and low ionization. In contrast, most $z>7$ galaxies observed with \jwst\ have high ratios of both R23 and O32, indicating intermediate metallicities ($\approx 0.1 Z_{\sun}$) and high ionization. \gal\ has the lowest known R23 of $z>7$ galaxies, indicating a much lower metallicity. In Figure \ref{fig:metal} we highlight the positions of two $z>8$ galaxies known to have low metallicities from the direct temperature method; ERO\_04590 (\citealt{Nakajima2023}, see also \citealt{Curti2023, Rhoads2023, Trump2023}) and EXCELS-63107 \citep{Cullen2025}. \gal\ has a significantly lower R23 ratio than both of these galaxies. 

We use R23-$Z$ relations calibrated from galaxies where direct electron temperatures have been measured from auroral lines. Using the relation of \cite{Nakajima2022} we obtain $12+{\rm log(O/H)}=6.88$ or 0.015\,$Z_{\odot}$. Using the \cite{Nakajima2022} relation for high EW(\Hb) galaxies would result in a small decrease to $12+{\rm log(O/H)}=6.85$ or 0.014\,$Z_{\odot}$.  If instead we use the relation of \cite{Curti2024} we get a similar $12+{\rm log(O/H)}=6.81$ or 0.013\,$Z_{\odot}$. We do not use the \cite{Sanders2024} relation since that relation is not well calibrated at low metallicities, but note that it would lead to substantially lower (0.3 dex) derived metallicity at $\sim0.01\,Z_{\odot}$.
We adopt the \cite{Nakajima2022} relation for high EW(\Hb) galaxies as our nominal metallicity calibration for this galaxy. The uncertainty on the metallicity includes their quoted uncertainty in the empirical calibration. However, we note that due to a paucity of such low-metallicity galaxies in the local Universe (only four galaxies in \citealt{Nakajima2022} have R23<2 and none of these have a high EW(\Hb) like \gal), there remains some uncertainty in the extrapolation of their calibration.

We do not have a detection of the \Oiiit\ line to derive the electron temperature and hence use the direct method of metallicity estimation. However, we note that some other similar extremely compact starbursts at these redshifts show very high electron temperatures, and in those cases the direct-method metallicity estimates are considerably lower than their R23-derived values \citep{Nakajima2023, Mowla2024, Cullen2025}. This is illustrated by comparing the two panels of Figure \ref{fig:metal} for our galaxy, ERO\_04590, Firefly, and EXCELS-63107. It is important to obtain deeper spectroscopy at higher resolution to determine if the metallicity of \gal\ should also be revised even lower. 

The mass-metallicity plane is shown in the right panel of Figure \ref{fig:metal}. \gal\ is a significant outlier from the distribution defined by $7<z<10$ galaxies from other \jwst\ studies. For context, we also plot on this figure the $z=6.6$ lensed star cluster, LAP-1, that has a strong-line metallicity estimate upper limit 0.5 dex lower than that of \gal. We use the plotted sample of $7<z<10$ galaxies from these studies to perform a fit of the mass-metallicity relation. \gal, EXCELS-63107, Firefly, and LAP-1 have been published specifically because of their low metallicities, so are excluded from the fit. Using the orthogonal distance regression method of \scipy, we find a best fit of 12+log(O/H)\,$ = 6.22 + 0.18\,log(M_* / M_{\odot})$. The slope of this relation lies in between those of \cite{Curti2024} and \cite{Nakajima2023}. The standard deviation of the galaxy metallicities with respect to the best fit line is 0.16 dex. \gal\ has a metallicity offset from the best fit line of $-0.77$ dex and is therefore a $4.7\sigma$ outlier with substantially lower metallicity than is typical for galaxies of similar mass at this epoch. 

\subsection{Lyman-$\alpha$ escape}
\label{sec:lyaesc}

One of the striking features of the spectrum of \gal\ is the strong \lya\ emission line with EW(Ly$\alpha)=63 \pm 9$\,\AA\ (Figure \ref{fig:spectrum}). The detection of this line suggests a considerable fraction of \lya\ photons are escaping from the galaxy and are transmitted through the circumgalactic medium and IGM. We calculate the fraction of escaping photons by considering the ratio between the observed fluxes of \lya\ and \Hb. We do not include any correction for dust, since the nebular dust extinction is not well constrained from the Balmer line ratios (Section \ref{sec:specfit}). We find a \lya\ escape fraction, f$_{\rm esc}$(Ly$\alpha$), of $0.21 \pm 0.05$ for Case B recombination, or $0.14 \pm 0.04$ for Case A. Taking Case B as our default value, this is the second-highest known \lya\ escape fraction at $z>8$ after the $z=8.28$ galaxy JADES-GN-z8-0-LA \citep{Tang2024, Witstok2025_3LAEs, Jones2025}. 

This significant escape fraction of \lya\ photons requires the galaxy to reside in an ionized bubble. The \lya\ and continuum fit in Figure \ref{fig:linefits} is consistent with a \lya\ line with zero intrinsic offset from the systemic redshift where the galaxy resides in a fully ionized bubble of size 2\,pMpc in an otherwise fully neutral IGM. With the limited S/N and resolution of the NIRSpec prism spectrum, we are unable to measure any intrinsic velocity offset or intrinsic line shape, for example if the line is already redshifted and asymmetric as it emerges from the galaxy ISM. An intrinsic redshift of the \lya\ line of many hundreds of km\,s$^{-1}$ would require a substantially smaller ionized bubble radius for similar \lya\ transmission \citep{Dijkstra2011}. However, \cite{Jones2025} show that the typical velocity offset for $z>6$ galaxies with such a high escape fraction as ours is only 200\,km\,s$^{-1}$, which would have a very small impact on the derived bubble radius \citep{MasonGronke2020}. It is important to confirm this assumption via future higher resolution spectroscopic observations of \lya. Assuming only a small velocity offset of $<200$ \,km\,s$^{-1}$, we can place a lower limit on the size of the ionized bubble by using the fact that the intrinsic \lya\ luminosity should not exceed that calculated from the \Hb\ luminosity, assuming Case B recombination. Under this assumption, we find a bubble size of $R_{\rm ion}>0.9$\,pMpc is required to not overproduce  \lya\ photons for the case of a fully neutral IGM. We find $R_{\rm ion}>0.7\,{\rm pMpc}$ if instead we assume the IGM has $x_{\rm HI}=0.7$.

A symmetric 1 to 2\,pMpc ionized bubble would be difficult to sustain for a galaxy of this UV luminosity.  Following the method of \cite{MasonGronke2020}, we estimate the size of the ionized bubble that the galaxy could generate based on balancing the ionizing photon output rate with recombinations. This method assumes the galaxy is initially embedded in a fully neutral IGM. Assuming the galaxy has been emitting ionizing photons for $\sim50$\,Myr (c.f. the {\texttt{Bagpipes}} estimated sSFR of $\sim10^{-8}\,{\rm yr}^{-1}$) with a relatively high Lyman continuum escape fraction of 0.5, the expected ionized bubble radius is $\approx 0.8$\,pMpc. This is lower than the estimated 1 to 2\,pMpc ionized bubble inferred by the high transmission of \lya.

\begin{deluxetable}{l l}
\tablewidth{120pt}
\tablecaption{\label{tab:galprop} Properties of \gal}
\startdata
& \\
CANUCS ID  &	2120090  \\
R.A. &	39.966109  \\
DEC &	$-1.594787$ \\
$z_{\rm spec}$ & $8.203 \pm 0.001$ \\
$\mu$\textdagger &	$8.0$  \\
$M_{*}$ &	$4.7_{-2.2}^{+2.9} \times 10^7 \,  M_{\sun}$\\
sSFR$_{\rm BAGPIPES}$ &	$1.1  \times 10^{-8}$ \,  yr$^{-1}$\\
SFR$_{\rm BAGPIPES}$  &	$0.54_{-0.26}^{+0.31} ~ M_{\sun}$ yr$^{-1}$\\
SFR$_{\rm H\beta}$    &	$0.44_{-0.05}^{+0.05} ~ M_{\sun}$ yr$^{-1}$\\
$M_{\rm UV}$ &	$-18.30 \pm 0.12$  \\
UV slope $\beta$ &	$-1.80 \pm 0.15$ \\
$r_{\rm hl}$ &	$38 ^{+3}_{-19} $\,pc\\
EW(Ly$\alpha$) & $63\pm9$\,\AA\\
EW(H$\beta$) & $225\pm50$\,\AA\\
EW(\Oiiib) & $297\pm63$\,\AA\\
Flux Ly$\alpha$ & $77 \pm 10\,\times 10^{-19}\,{\rm erg \, s}^{-1} \, {\rm cm}^{-2}$\\
Flux \Oii$\ddagger$ & $<2.3\,\times 10^{-19}\,{\rm erg \, s}^{-1} \, {\rm cm}^{-2}$\\
Flux H$\delta$ & $3.0 \pm 1.3\,\times 10^{-19}\,{\rm erg \, s}^{-1} \, {\rm cm}^{-2}$ \\
Flux H$\gamma$ & $3.9 \pm 1.4\,\times 10^{-19}\,{\rm erg \, s}^{-1} \, {\rm cm}^{-2}$ \\
Flux \Oiiit$\ddagger$ & $<2.0\,\times 10^{-19}\,{\rm erg \, s}^{-1} \, {\rm cm}^{-2}$ \\
Flux H$\beta$ & $15.6 \pm 1.8\,\times 10^{-19}\,{\rm erg \, s}^{-1} \, {\rm cm}^{-2}$ \\
Flux \Oiiia & $6.9 \pm 0.6\,\times 10^{-19}\,{\rm erg \, s}^{-1} \, {\rm cm}^{-2}$ \\
Flux \Oiiib & $20.6 \pm 0.6\,\times 10^{-19}\,{\rm erg \, s}^{-1} \, {\rm cm}^{-2}$ \\
R23 & $1.76 \pm 0.23$\\
R3 & $1.32 \pm 0.15$\\
O32 & $>8.80$\\
12+log(O/H)   & $6.85\pm 0.16$ (0.014\,$Z_{\sun}$)\\
$f_{\rm esc}$(Ly$\alpha$) case B & $0.21 \pm 0.05$ \\
$f_{\rm esc}$(Ly$\alpha$) case A  & $0.15 \pm 0.04$ \\
\xiion & $25.56 \pm 0.08$\, \xiunit \\
\hline
\enddata
\tablenotetext{}{\textdagger For a discussion of the uncertainty on the lensing magnification $\mu$ see Section \ref{sec:lensmod}. Quoted values of $M_{*}$, SFR, and $M_{\rm UV}$ are corrected for the best fit $\mu = 8.0$, but their uncertainties do not include the additional uncertainty in magnification. The uncertainty on the half-light radius does include the uncertainty in magnification.\\ $\ddagger$ Undetected emission lines are quoted with 2$\sigma$ upper limits to their fluxes.}
\end{deluxetable}

A possible solution to this inconsistency is that \gal\ resides in an overdense environment where other galaxies with bursts of star formation at different epochs have combined to produce a larger ionized region. Support for this notion comes from an analysis of spectroscopic redshifts in the field. There are four galaxies with $z_{\rm spec}>8$ in the Abell 370 field measured from CANUCS NIRSpec spectroscopy \citep{Sarrouh2025arXiv}. Three of these (including \gal) have redshifts equal to within $\delta z = 0.01$. The other two galaxies in this redshift interval are CANUCS-2110006 and CANUCS-2111943 with redshifts $z = 8.199 \pm 0.001$ and $8.198 \pm 0.004$, respectively. Of relevance to the environment of \gal, these two other galaxies are located at projected distances of only 0.09 and 0.12 \,pMpc from \gal, close enough to be contained within the same ionized bubble. 
We hypothesize that these galaxies, potentially combined with other galaxies in this overdense region, have aided the generation of a relatively large ionized bubble.  Further details of these galaxies and the implications for the ionization of the IGM in this region will be presented in Harshan et al. (in preparation).

\subsection{Star formation density}
\label{sec:sfrrho}
With the aid of gravitational lensing, we have strong constraints on the size of \gal, showing it to be extremely compact. From the half-light radius for a symmetric source of $r_{\rm hl} = 38 ^{+3}_{-19} $\,pc and taking half of SFR$_{\rm H\beta}$, we calculate the star-formation rate density, $\Sigma_{SFR}$, within this region to be in the range $50 - 100\,M_{\sun}$\,yr$^{-1}$\,kpc$^{-2}$, depending on the assumed lensing magnification. 
This density is significantly higher than local Green Pea galaxies \citep{Yang2017} and 10 times greater than the median for $7<z<9$ galaxies \citep{Morishita2024sizes}. The SFR density of \gal\ is comparable to other strong \lya-emitters at Cosmic Noon \citep{Kim2025arXiv} and in the Epoch of Reionization (COLA-1; \citealt{Matthee2018} and JADES-GN-z8-0-LA; \citealt{Witstok2025_3LAEs}). The SFR densities of these galaxies are high, considering the Eddington limit for an optically thick star-forming disk is $\sim 1000 \,M_{\sun}$\,yr$^{-1}$\,kpc$^{-2}$ \citep{Thompson2005}. We can speculate that both the high \lya\ escape fraction and low metallicity may be related to the extreme SFR density (see \citealt{Pucha2022}). 

\section{Discussion}

The discovery of \gal\ with $z=8.203\pm 0.001$ is important for a few reasons: it is the first $z>7$ galaxy found to have a metallicity approaching $1\%$ solar; its extreme properties, including its compact size ($r_{\rm hl} \sim 40\,{\rm pc}$), young age ($\lesssim 20\,{\rm Myr}$) and high star formation surface density ($\Sigma_{SFR}\sim 50 - 100\,M_{\sun}$\,yr$^{-1}$\,kpc$^{-2}$) indicate a burst of star formation under extreme conditions; and its two close companions suggest that it resides in an ionized bubble at an epoch when the Universe was still predominantly neutral. Taken together, these factors provide valuable constraints to test our theories about the mechanisms driving reionization and galaxy evolution at $z>8$.

{\it Low metallicity}. The exceptionally low metallicity of \gal\ is particularly noteworthy, as it represents one of the first z > 7 galaxies found with metallicity approaching 1\% solar. This finding aligns with theoretical expectations from cosmological simulations like ASTRAEUS that predict the existence of very low metallicity systems at these redshifts, especially in lower-mass halos \citep{Hutter2021,Legrand2022}. However, the relative scarcity of such low-metallicity objects in current observations (Figure\,\ref{fig:metal}) is puzzling. One possible explanation lies in the galaxy's extreme compactness - its $\approx 40$\,pc half-light radius allows us to probe star formation on scales comparable to individual star-forming regions, potentially providing a cleaner window into pristine gas conditions than in more extended systems where metallicity measurements are averaged over larger volumes \citep{James2020}. It is notable that the lowest metallicity object known in the reionization epoch is the strongly-lensed $z=6.6$ compact object LAP-1 \citep{Vanzella2023} that has a mass equivalent to a single star cluster.

{\it Formation in an overdense region}. The galaxy's environment appears to play a crucial role in its evolution. The presence of two other galaxies within 120 kpc at nearly identical redshifts suggests \gal\ resides in an overdense region, possibly near a cosmic filament. This environment could explain several of its unique properties. 

Cosmological simulations by \citet{Ceverino2016} demonstrate that gas inflow from the cosmic web can create star-forming regions with metallicities $\sim 0.3$ dex lower than the surrounding ISM, particularly at high redshift. The specific combination of properties we observe - stellar mass ($\sim 5 \times 10^7\,M_{\sun}$), high sSFR ($\sim 10^{-8}$\,yr$^{-1}$), and low metallicity ($\sim 1\%\, Z_\sun$) - aligns with theoretical predictions from \citet{Marszewski2024} who find in FIRE-2 simulations show that our galaxy's position in the mass-metallicity plane is consistent with their models when accounting for gas inflows and outflows. Additionally, \citet{Yajima2015} provide specific theoretical scaling relations between stellar mass, gas fraction, metallicity, and SFR at high redshift which show that low metallicities in galaxies like \gal\ can arise from a dilution due to the inflow of pristine gas, rather than a newly formed galaxy experiencing its first epoch of star formation.  

The environmental context strengthens this interpretation. \citet{Chiang2017} show that protoclusters have a significant volume filling fraction at $z\sim7$ and exhibit elevated SFRs, with \citet{Harshan2023} finding that galaxy properties at high-z are significantly correlated with their environment. This aligns with our observation of multiple galaxies in close proximity to \gal\ at $z\sim8$, suggesting an overdense environment. Within the framework of the gas-regulator model \citep{Lilly2013}, such an environment would enhance cold gas accretion along cosmic filaments, providing both the fuel for the observed star formation and explaining the low metallicity through dilution of the existing metal content of the ISM. Given the substantial stellar mass already built up in this galaxy, gas dilution is a more likely explanation for the low metallicity than the galaxy being in a phase of forming its first generation of stars.

{\it Ly$\alpha$ emission, UV slope and extreme star formation}. The strong \lya\ emission (EW(Ly$\alpha$) = 63 Å) and high escape fraction ($f_{\rm esc}$(Ly$\alpha$) = 0.21) are remarkable given the highly neutral IGM expected at $z = 8.2$. Recent work by \cite{Witten2024} demonstrates that tidal interactions and bursty star formation due to merging galaxies can create channels for \lya\ escape. However, unlike the sample of LAEs of \cite{Witten2024}, \gal\ does not have a merging, close-companion galaxy. The extremely high SFR density suggests that \gal\ may represent a fundamental mode of early galaxy evolution where star formation proceeds at maximal efficiency. This connects with theoretical work showing that such compact, intense star formation may be crucial for reionization, as the mechanical energy from intense star formation can more efficiently create channels for ionizing photon escape through stellar feedback \citep{Trebitsch2017, Menon2024}. 

The relatively red UV slope ($\beta=-1.8$) presents an interesting puzzle given the low metallicity and strong \lya\ emission. A similar UV slope is seen in ERO\_04590 ($\beta=-1.7$; \citealt{Fujimoto2024}) and attributed to moderate dust reddening. These slopes are in stark contrast to EXCELS-63107, which has a very blue UV slope ($\beta=-3.3$; \citealt{Cullen2025}). While the red slope of \gal\ could indicate dust enrichment, which seems at odds with the system's other properties, an alternative explanation may lie in the contribution of nebular continuum emission, consistent with the very high \Hb\ equivalent width. However, nebular continuum is expected to have a somewhat bluer slope than seen in \gal, so this aspect remains uncertain. There are also models of massive Population III stars that predict relatively red colors due to reprocessing of radiation in inflated stellar envelopes \citep{Woods2021}. Further work is required to model and interpret the complex SEDs of these galaxies during the Epoch of Reionization.



Taken together, \gal\ may represent a crucial but short-lived phase in early galaxy evolution where pristine gas accretion, extreme star formation, and efficient feedback processes combine to produce unique observational signatures. Its properties suggest that it may be capturing the moment when one of the Universe's earliest galaxies is beginning to enrich its surroundings and contribute to cosmic reionization.

\section{Conclusions}

We have presented JWST imaging and spectroscopy of \gal\ showing that it is an ultracompact, very low metallicity \lya\ emitter at $z=8.203\pm 0.001$. Our key findings are summarized below:

\begin{itemize}

\item{The R23 ratio of $1.76 \pm 0.23$ is the lowest of any known $z>7$ galaxies, where \Hb\ is not enhanced due to the presence of an AGN broad line. The gas-phase metallicity derived from R23 is 12+log(O/H) $= 6.85\pm 0.16$ or 1.4\% solar. The galaxy is a $4.7\sigma$ outlier from the $z>7$ mass-metallicity relation.}

\item{Strong \lya\ emission is visible in the low-resolution NIRSpec spectrum with EW(Ly$\alpha) = 63\pm9$\,\AA. The \lya\ escape fraction is estimated to be $\sim 20\%$, the second highest known for galaxies at this epoch. The finding of two companion galaxies at the same spectroscopic redshift $\approx 100$\,kpc away suggests an overdensity of galaxies that may have reionized this region and allowed the transmission of \lya\ photons.}

\item{The galaxy has a half-light radius of only $\approx 40$\,pc and a star formation density $\Sigma_{SFR}\sim 50 - 100\,M_{\sun}$\,yr$^{-1}$\,kpc$^{-2}$. The combination of high SFR density and low metallicity suggests that this galaxy is going through a rare evolutionary stage of rapid cold gas accretion from the cosmic web. The low metallicity is therefore more likely to be due to a recent inflow of low-metallicity gas rather than a young galaxy forming its first stellar populations.}

\end{itemize}

Follow-up spectroscopic observations and observations of similar systems will be crucial for testing our interpretation and refining theoretical models of reionization and galaxy formation in the early Universe.

\section*{Acknowledgements}

Thanks to Seiji Fujimoto for useful discussions and the anonymous reviewer for their interesting and constructive suggestions.
This research was enabled by grant 18JWST-GTO1 from the Canadian Space Agency and funding from the Natural Sciences and Engineering Research Council of Canada.
Y.A. is supported by a Research Fellowship for Young Scientists from the Japan Society of the Promotion of Science (JSPS).
M.B., J.J., N.M., A.H. and G.R. acknowledge support from the ERC Grant FIRSTLIGHT and from the Slovenian national research agency ARRS through grants N1-0238 and P1-0188.
This research used the Canadian Advanced Network For Astronomy Research (CANFAR) operated in partnership by the Canadian Astronomy Data Centre and The Digital Research Alliance of Canada with support from the National Research Council of Canada, the Canadian Space Agency, CANARIE and the Canadian Foundation for Innovation.

\section*{Data Availability}

Data presented in this paper will be made available upon request. The CANUCS DOI is \url{doi.org/10.17909/ph4n-6n76}.

\facilities{\jwst\ (NIRCam, NIRISS, NIRSpec)}

\bibliographystyle{aasjournal}
\bibliography{references} 


\end{document}